\documentclass[a4paper,11pt]{article}
\usepackage[english]{babel}
\usepackage{braket}
\usepackage{amsmath,amsfonts,graphicx,color,cite}
\usepackage{ulem}

\newcommand{\no}{\nonumber}
\newcommand{\beeq}{\begin{equation}}
\newcommand{\eeq}{\end{equation}}
\newcommand{\bee}{\begin{eqnarray}}
\newcommand{\ee}{\end{eqnarray}}
\newcommand{\been}{\begin{eqnarray*}}
\newcommand{\een}{\end{eqnarray*}}

\makeatletter\@addtoreset{equation}{section}
\makeatother

\begin{document}
\begin{titlepage}
\begin{flushright}
TIT/HEP-666\\
June, 2018
\end{flushright}
\vspace{0.5cm}
\begin{center}
{\Large \bf
T-duality to
Scattering Amplitude and Wilson Loop in
 Non-commutative Super Yang-Mills Theory
}
\lineskip .75em
\vskip 1.5cm
{\large  Song He$^{a}$~\footnote{hesong17@gmail.com}~,
Hongfei Shu$^{b}$~\footnote{h.shu@th.phys.titech.ac.jp}
}
\date{}

\vskip 1.0cm
\vspace{-10mm}
\begin{center}
{\it
$^{a}$Max Planck Institute for Gravitational Physics (Albert Einstein Institute),\\
Am M\"uhlenberg 1, 14476 Golm, Germany\\\vspace{2mm}
$^{b}$ Department of Physics, Tokyo Institute of Technology, Tokyo, 152-8551, Japan\\\vspace{2mm}
}
\vspace{10mm}
\end{center}

\vskip 3.0em
\end{center}

\begin{abstract}
{We first perform bosonic T-duality transformation on one of the marginal TsT (T-duality, shift, T-duality)-deformed $AdS_5\times S_5$ spacetime, which corresponds to 4D $\mathcal{N}=4$ non-commutative super Yang-Mills theory (NCSYM). We then construct the solution to Killing spinor equations of the resulting background, and perform the fermionic 
T-duality transformation. The final dual geometry becomes the usual $AdS_5\times S_5$ spacetime but with a constant 
NS-NS B-field depending on the non-commutative parameter. As applications, 
we study the gluon scattering amplitude and open string (Wilson loop) solution in the TsT-deformed $AdS_5\times S_5$ spacetime, which are dual to the null polygon Wilson loop and the folded string solution respectively in the final dual geometry.}

\end{abstract}
\end{titlepage}
\baselineskip=0.7cm
\tableofcontents

\section{Introduction}
In the past decades, many perturbative and non-perturbative results have been achieved in the planar limit of ${\cal N}=4$ super Yang-Mills theory (SYM). See \cite{Beisert:2010jr} for reviews about the aspects of spectrum, scattering amplitude and Wilson loop. One of the significant achievements about scattering amplitude and Wilson loop is made in \cite{Alday:2007hr}. It has been shown that 
$n$-gluon scattering amplitude at strong coupling in ${\cal N}=4$ SYM can be calculated from the minimal area of the surface ending on a null $n$-polygonal Wilson loop at the boundary of AdS space. This is based on the self-duality of IIB string theory under a certain combination of bosonic and fermionic T-duality transformations in $AdS_5\times S^5$ spacetime \cite{Beisert:2008iq,Berkovits:2008ic}, 
which also explains that the existence of the dual superconformal symmetry
\cite{Berkovits:2008ic}. There are many ways to deform ${\cal N}=4$ SYM. Deformed field theories arising from a new definition of the product of fields in the Lagrangian, provide an interesting generalization of the gauge/gravity correspondence \cite{Maldacena:1997re, Gubser:1998bc,  Witten:1998qj}. This is also due to the fact that, on the string theory side, there is a systematic
procedure called the ``TsT transformation'' (T-duality, shift, T-duality) \cite{Lunin:2005jy,Frolov:2005dj}. The most well-known examples are non-commutative deformation \cite{Hashimoto:1999ut,Maldacena:1999mh,Alishahiha:1999ci}, $\beta$-deformation \cite{Leigh:1995ep} and ``dipole deformation'' \cite{Bergman:2000cw,Dasgupta:2000ry,Bergman:2001rw,Dasgupta:2001zu} of $AdS_5\times S^5$ spacetime. A good summary can be found in \cite{Imeroni:2008cr}, see also the reference therein.

We mainly focus on the holography dual of scattering amplitudes and Wilson loop operators in ${\cal N}=4$ non-commutative super Yang-Mills theory (NCSYM). In \cite{Sever:2009xu}, the author has holographically studied the planar gluon scattering amplitudes in terms of scattering amplitude/Wilson loop duality \footnote{{See \cite{Georgiou:2010eq} for a special scattering amplitudes amplitude in the finite temperature regime of non-commutative Tang-mills theory.}}. However, the fermionic T-duality has not been considered, which leads to non-constant dilaton and complex field strength.

On the other hand, many works about the fermionic T-duality transform and the scattering amplitude/Wilson loop duality have been done in the past years \cite{Arutyunov:2008if, Stefanski:2008ik, Gomis:2008jt, Godazgar:2010ph, Bakhmatov:2011aa, Bianchi:2011dg, Adam:2010hh, Bakhmatov:2010fp, Sorokin:2011mj, Dekel:2011qw, OColgain:2012si, Abbott:2015mla, Bakhmatov:2015wdr, Abbott:2015ava, Colgain:2016gdj}%
\footnote{These works \cite{Arutyunov:2008if, Stefanski:2008ik, Gomis:2008jt, Godazgar:2010ph, Bakhmatov:2011aa, Bianchi:2011dg, Adam:2010hh, Bakhmatov:2010fp, Sorokin:2011mj} are focus on the ABJM theory and the other works \cite{Dekel:2011qw, OColgain:2012si, Abbott:2015mla, Bakhmatov:2015wdr, Abbott:2015ava, Colgain:2016gdj} study the ${\cal N}=4$ SYM theory.}. 
One of the purposes of the present work is to present the construction of fermionic T-duality transformation of the NCAdS background to cancel the non-constant dilaton and the complex field strength, and corresponding string solutions in the final dual background in more details which are expected to dual to scattering amplitude. Following the notaton in \cite{Sever:2009xu}, we have called the gravity background corresponding to NCSYM as NCAdS spacetime, whose definition will be shown in (\ref{eq:NCAdS-background}).
More precisely we first perform bosonic T-duality transformation on the NCAdS background, and construct the solution to Killing spinor equations of the resulting background. We then perform the fermionic 
T-duality transformation, and find the final dual background which is expected to be equivalent with the NCAdS background. 
In the final dual background, we construct the solution with proper boundary conditions which is expected to be holographic dual of gluon amplitude in NCSYM.
Further, a motivation \cite{Kruczenski:2012aw} to explore the relation between closed and open strings in AdS leads us to construct the open string solution (Wilson loop) ending at the boundary of NCAdS, which dual to closed string in the final dual background. This would be important in the study of the closed and open strings relation and the corresponding observables in NCSYM.

The layout of this paper is as follows.
In sec.\ref{sec:TsT}, we first review the TsT transformation of $AdS_5\times S^5$ spacetime (NCAdS spacetime) which corresponds to the non-commutative ${\cal N}=4$ Super Yang-Mills theory. We then perform certain bosonic and fermionic T-dual transformations on the NCAdS background, and obtain a simplified gravity background. In sec.\ref{sec:Amplitude-NC}, we study the scattering amplitudes in NCSYM by using the solutions in the simplified gravity background. In sec.\ref{sec:Open string Solution in NCAdS}, we construct the open string solution in the NCAdS background, which is dual to the folded string solution in the simplified gravity background. Finally, we devote to the conclusions and discussions and also mention the future problems. In appendices, we would like to list some techniques and Elliptic functions which are very useful in our analysis.

\section{TsT deformed $AdS_5\times S^5$ spacetime and their T-duality transformations}\label{sec:TsT}
The gravity dual of non-commutative gauge theory in \cite{Hashimoto:1999ut,Maldacena:1999mh} can be generated from the $AdS_5\times S^5$ spacetime by using TsT deformation $(x^1,x^2)_\gamma$, which stands the T-dualizing $x^1 \to {x}^{1}_t$, shift $x^2$ by $x^2+\gamma {x}^1_t$, then T-dualizing back ${x}^{1}_t \to x^1$ \cite{Lunin:2005jy}, 
where ${x}_t^{\mu}$ $(\mu=0,1,2,3)$ is the coordinate T-dual to $x^{\mu}$ and $\gamma$ is the constant deformation parameter.

By applying the $(x^1,x^2)_\gamma$ TsT-deformation to $AdS_{5}\times S^{5}$ spacetime, 
one obtains the following background: 
\begin{eqnarray}\label{eq:NCAdS-background}
	ds^{2}&=&\frac{R^{2}}{r^{2}}\left(-dx_{0}^{2}+dx_{3}^{2}+dr^{2}\right)
+\frac{R^{2}}{r^{2}}\frac{1}{1+\gamma^{2}\frac{R^{4}}{r^{4}}}
\left(dx_{1}^{2}+dx_{2}^{2}\right)+R^{2}ds_{S^{5}}^{2}\no\\
	B&=&\frac{\gamma R^{4}}{r^{4}}\frac{1}{1+\gamma^{2}\frac{R^{4}}{r^{4}}}
dx^{1}\wedge dx^{2},\quad
        \phi=-\frac{1}{2}\log\left(1+\gamma^{2}\frac{R^{4}}{r^{4}}\right), \\
 F_{1}&=&0, \quad
    	F_{3}=
-4\gamma\frac{R^{4}}{r^{5}}dx^{0}\wedge dx^{3}\wedge dr,
\no\\
	F_{5}&=&-4R^{4}\left(\frac{1}{1+\gamma^{2}\frac{R^{4}}{r^{4}}}
\frac{1}{r^{5}}dx^{0}\wedge dx^{1}\wedge dx^{2}\wedge dx^{3}\wedge dr+\omega_{S^5}\right), \no
\end{eqnarray}
where $ds^{2}$ is the metric of the NCAdS background, $B$ is the NS-NS B-field, $\phi$ is dilaton and ${F}_{1}$, ${F}_{3}$, 
${F}_{5}$ are the R-R field strengths.

The dual gauge theory of this background 
\eqref{eq:NCAdS-background} is defined on the noncommutative spacetime with noncommutativity parameter
$[x^{1},x^{2}]=i\theta^{12}$. The constant noncommutativity parameter 
$\theta^{12}$\cite{Seiberg:1999vs} is associated with the TsT-deformation parameter $\gamma$ as \cite{Hashimoto:1999ut}
\bee
\theta^{12}=2\pi\alpha'\left(\left.B_{12}\right|_{r\to 0}\right)^{-1}
=2\pi\alpha'\gamma.
\ee
Then the so called Seiberg-Witten $\alpha'\to 0$ limit has to be taken with 
$\alpha'\gamma$ fixed. 

For large $r$, background (\ref{eq:NCAdS-background}) reduces to the original $AdS_5$$\times S^5$ spacetime. Since the large $r$ region corresponds to the IR regime of the gauge theory, one expects that the non-commutative super Yang-Mills theory reduces to the original $N=4$ super Yang-Mills theory at IR regime (long distance). We will study the gluon scattering amplitude at IR regime in section \ref{sec:Amplitude-NC}.  The background (\ref{eq:NCAdS-background}) has boundary at $r=0$. Since $G_{11}, G_{22}\propto \frac{r^2}{R^2}$ in the boundary of background (\ref{eq:NCAdS-background}), the physical size of $x_1$ and $x_2$ directions shrink \cite{Maldacena:1999mh}.

\subsection{Bosonic T-duality transform along non-radial directions}
\label{subsec:Bosonic_T-dual}
The background (\ref{eq:NCAdS-background}) is invariant under the the shift isometry $x^\mu\to x^\mu+c$, where $\mu=0,1,2,3$, $c$ is a constant. We then
perform 
T-duality transform along 
$x^1,x^2,x^3,x^0$ in turn by following the Buscher rule in appendix \ref{sec:BT-dual}. The resulting background becomes 
\begin{eqnarray}\label{eq:TsTT4}
	ds^{\prime}{}^{2}&=&\frac{r^{2}}{R^{2}}
\left(-(d{x}^{\prime 0})^{2}+(d{x}^{\prime 1})^{2}+(d{x}^{\prime 2})^{2}+(d{x}^{\prime 3})^{2}\right)
+\frac{R^{2}}{r^{2}}dr^{2}+R^{2}ds_{S^{5}}^{2},
    \no\\
	B'&=&-\gamma d{x}^{\prime 1}\wedge d{x}^{\prime 2},
    \quad
\phi'=\log\left(\frac{r^{4}}{R^{4}}\right),
    \no\\
{F}'_{1}&=&-4i\frac{R^{4}}{r^{5}}dr, \quad
{F}'_{3}={F}'_{5}=0, 
\end{eqnarray}
This background is same as the one obtained from the T-dual transformation of original $AdS_5$ along $x^1,x^2,x^3,x^0$ directions but with a constant $B$-field. 
We note that
the factor $i$ has appeared in ${F}'_1$ due to T-dualizing along the 
time direction $x^0$ \cite{Hull:1998vg}. To cancel the dilaton and the complex field strength, we perform fermionic T-dual transformation on the background (\ref{eq:TsTT4}). 

\subsection{Fermionic T-duality transform}
In this subsection, we perform the fermionic T-duality on the background (\ref{eq:TsTT4})
\footnote{We follow the notation summarized in \cite{Bakhmatov:2009be} 
.}. Given Killing spinors $(\epsilon_I,\hat{\epsilon}_I)$, the fermionic 
T-duality transform is generated as \cite{Bakhmatov:2009be}
\begin{eqnarray}
0&=&\epsilon_{I}\gamma^{M}\epsilon_{J}+\hat{\epsilon}_{I}\gamma^{M}\hat{\epsilon}_{J}
\label{eq:abelian}\\
\partial_{M}C_{IJ}&=&2i\epsilon_{I}\gamma_{M}\epsilon_{J}
\label{eq:FT-rule}\label{eq:FT-C-eps}\\
\tilde{\phi}&=&\phi^\prime+\frac{1}{2}\mbox{Tr}(\log C)\label{eq:FT-phi}\label{eq:FT-phi-C}\\
\frac{i}{16}e^{{\tilde{\phi}}}{\tilde{F}}^{\alpha\hat{\beta}}&=&\frac{i}{16}e^{\phi^\prime}F^{\prime\alpha\hat{\beta}}-\epsilon_{I}^{\alpha}\hat{\epsilon}_{J}^{\hat{\beta}}(C^{-1})_{IJ}\label{eq:FT-F},
\end{eqnarray}
where the indices $I$, $J$ are the labels of the different Killing spinors, 
and $F^{\alpha\hat{\beta}}$ are the R-R fields in bispinor form 
\bee
F^{\alpha\hat{\beta}}=(\gamma^{M})^{\alpha\hat{\beta}}F_{M}+\frac{1}{3!}(\gamma^{MNP})^{\alpha\hat{\beta}}F_{MNP}+\frac{1}{2}\frac{1}{5!}(\gamma^{MNPQR})^{\alpha\hat{\beta}}F_{MNPQR},\quad M=0,1,\cdots,9.
\ee
$\tilde{F}^{\alpha\hat{\beta}}$ and $F^{\prime\alpha\hat{\beta}}$ are also 
defined in a similar way. Note that the metric and the B-field do not transform under the above fermionic T-duality. The Killing spinors are not arbitrary. More precisely, \eqref{eq:abelian} is imposed to ensure the fermionic isometries, i.e. $\{\epsilon^{I}Q_{I},\hat{\epsilon}^{\hat{I}}Q_{\hat{I}}\}=0$ where $Q$ and $\hat{Q}$ are the supercharges in the supersymmetry algebra.
Relaxing
\eqref{eq:abelian}, fermionic T-duality will lead to field configurations that
are not supergravity solutions. The second equation determines the matrix $C$.
In order to do that, we have to find 
the unbroken supersymmetry in 10D IIB supergravity, 
which is generated by the spinor parameters 
$(\epsilon^{\alpha},\hat{\epsilon}^{\hat{\alpha}})$ 
$(\alpha,\hat{\alpha}=1,\ldots,16)$. 
Then the parameters must satisfy the following equations from 
the supersymmetry transformation of 
two gravitini $\psi_{M}$, $\hat{\psi}_{M}$ $(M=0,\ldots,9)$ and two 
dilatini $\lambda$, $\hat{\lambda}$: 
\bee
\delta\psi_{M}&=&{e_{\hat{M}}}^{M}\nabla_{M}\epsilon-\frac{e^{\phi^\prime}}{8}\gamma^{\hat{N}}F'_{\hat{N}}\gamma_{\hat{M}}\hat{\epsilon}=0,\label{eq:gravitino1}\\
\delta\hat{\psi}_{\hat{M}}&=&{e_{\hat{M}}}^{M}\nabla_{M}\hat{\epsilon}+\frac{e^{\phi^\prime}}{8}\gamma^{\hat{N}}F'_{\hat{N}}\gamma_{\hat{M}}\epsilon=0,\label{eq:gravitino2}\\
\delta\lambda&=&{e_{\hat{M}}}^{M}\gamma^{\hat{M}}\partial_{{M}}\phi^\prime\epsilon+e^{\phi^\prime}\gamma^{\hat{M}}F^\prime_{\hat{M}}\hat{\epsilon}=0,\label{eq:dilatino1}\\
\delta\hat{\lambda}&=&{e_{\hat{M}}}^{M}\gamma^{\hat{M}}\partial_{M}\phi^\prime\hat{\epsilon}-e^{\phi^\prime}\gamma^{\hat{M}}F'_{\hat{M}}\epsilon=0,\label{eq:dilatino2}
\ee
where $\delta$ is the supersymmetry variation, $\nabla_{M}$ is 
the covariant derivative and $\gamma^{\hat{M}}$ is the 10D gamma matrices. $\hat{M}$ is the coordinate of the flat space, and ${e_{\hat{M}}}^N$ is the the vielbein. From the dilatino equations \eqref{eq:dilatino1} and \eqref{eq:dilatino2}, 
$\epsilon$ and $\hat{\epsilon}$ are related with
\bee
\hat{\epsilon}&=&-i\epsilon.
\label{eq:complex}
\ee
Although the supersymmetric parameters in type IIB supergravity are real Majorana-Weyl spinors, 
an imaginary
unit appears because of the complexified RR one form (\ref{eq:TsTT4}).
Therefore, (\ref{eq:abelian}) is satisfied automatically. Substituting \eqref{eq:complex} the gravitino equations are simplified as 
\bee
{e_{\hat{M}}}^{M}\nabla_{M}\epsilon+i\frac{e^{\phi^\prime}}{8}\gamma^{\hat{4}}F'_{\hat{4}}\gamma_{\hat{M}}
\epsilon=0.
\label{eq:Killingeq}
\ee
We can find that the $M=x^{0,1,2,3}$ part of \eqref{eq:Killingeq} is trivial. 
To solve the remaining 6D part ($R\times S^5$), it is convenient to rewrite the coordinate as $y^s$ ($s=1,2,\cdots,6$) with $|y|=r$. We decompose SO(9,1) spinor index $\alpha$ into 
$(a' j', \dot{a}'j')$ as in \cite{Berkovits:2008ic}, 
where $a',\ \dot{a}'=1,2$ are the SO(3,1) spinor indices and $j'=1,\ldots,4$ 
is the SO(5) spinor index. 
$(\sigma^{r})_{j'k'}$ $(r=1,\ldots,6)$ are the 6D Pauli matrices. 
Then \eqref{eq:Killingeq} is solved as
\bee
\epsilon_{aj}^{b'l'}=\sqrt{\frac{r}{R}}\delta_a^{b'}M_j^{l'}(y),\quad
\epsilon_{aj}^{\dot{b}'l'}=0,\quad \hat{\epsilon}_{aj}^{b'l'}=-i\epsilon_{aj}^{b'l'}
\label{eq:Killing}
\ee
where $M_j^{l'}(y)$ is the SU(4)/SO(5) matrix rotating the point 
$(0,0,0,0,0,1)$ on $S^5$ to the point $(y_1,y_2,y_3,y_4,y_5,y_6)/r$.
The unprimed indices $a$, $j$ are regarded as the label of different Killing spinors corresponding to the label $I$ in (\ref{eq:FT-C-eps}). 
The main difference between our Killing spinor and the chosen one in \cite{Berkovits:2008ic} is (\ref{eq:complex}). We will show the details of the Killing spinors of original $AdS_5\times S^5$ spacetime used in \cite{Berkovits:2008ic} in appendix \ref{sec:KSE-two}. Then one finds 
\bee
C_{aj~bk}=2i\epsilon_{ab}\sigma_{jk}^{r}y_{r},\quad(C^{-1})^{aj~bk}=-\frac{i}{2}\epsilon^{ab}(\sigma^{r})^{jk}\frac{y_{r}}{r^{2}}.
\ee
 To determine the transformation of field strength strength, we use
\bee
\epsilon_{aj}^{a'j'}(C^{-1})^{aj~bk}\hat{\epsilon}_{bk}^{b'k'}=-\frac{1}{2R}\epsilon^{a'b'}(\sigma^{6})^{j'k'}=\frac{i}{2R}(\gamma_{\hat{0}\hat{1}\hat{2}\hat{3}\hat{4}})^{a'j'~b'k'}.
\ee
Writing it in the terms of
projection operator $\frac{1}{2}\big((\gamma_{\hat{0}\hat{1}\hat{2}\hat{3}}-i)\gamma_{\hat{4}}\big)^{\alpha\hat{\beta}},$
one finds
\begin{align}
e^{\tilde{\phi}}\tilde{F}^{\alpha\hat{\beta}} 
 & =-\gamma^{\hat{4}}i\frac{4}{R}-\frac{4}{R}(\gamma_{\hat{0}\hat{1}\hat{2}\hat{3}\hat{4}}-i\gamma_{\hat{4}})^{\alpha\hat{\beta}}=-\frac{4}{R}(\gamma_{\hat{0}\hat{1}\hat{2}\hat{3}\hat{4}})^{\alpha\hat{\beta}}.
\end{align}
Furthermore, using (\ref{eq:FT-phi-C}) one finds the dilaton vanishes $\tilde{\phi}=0$.
Introducing $z=\frac{R^{2}}{r}$ and $\tilde{x}^\mu=x^{\prime \mu}$, we then obtain the dual background as 
\begin{eqnarray}\label{eq:TsTT4FT}
	d\tilde{s}^{2}&=&\frac{R^{2}}{z^{2}}
\left(-(d\tilde{x}^{0})^{2}+(d\tilde{x}^{1})^{2}+(d\tilde{x}^{2})^{2}+(d\tilde{x}^{3})^{2}+dz^{2}\right)+R^{2}ds_{S^{5}}^{2},\no\\
        \tilde{B}&=&-\gamma d\tilde{x}^{1}\wedge d\tilde{x}^{2},\quad\tilde{\phi}=0\no\\
        \tilde{F}_{1}&=&\tilde{F}_{3}=0,\quad
        \tilde{F}_{5}=-4R^{4}\left(\frac{1}{z^{5}}d\tilde{x}^{0}\wedge d\tilde{x}^{1}\wedge d\tilde{x}^{2}\wedge d\tilde{x}^{3}\wedge dz+\omega_{S^5}\right).
\end{eqnarray}
This is the usual $AdS_5\times S^5$ background but with the constant 
B-field. 
One may suspect the above fermionic T-duality transform because we have 
performed it in complex background. In order to resolve this issue, we consider the whole duality transform 
in reversed order. First, starting from the background \eqref{eq:TsTT4FT},
we perform the fermionic T-duality transformation. Since the background is real and the
constant B-field does not contribute to the Killing spinor equations, 
we can use the same spinors $(\epsilon, \hat{\epsilon})$ 
as the ones in \cite{Berkovits:2008ic}
\bee\label{eq:KS-in-BM}
\epsilon_{aj}^{b'l'}=\sqrt{\frac{r}{R}}\delta_a^{b'}M_j^{l'}(y),\quad
\epsilon_{aj}^{\dot{b}'l'}=0,\quad \hat{\epsilon}_{aj}^{b'l'}=i\epsilon_{aj}^{b'l'}.
\ee
Then using transformation rules (\ref{eq:FT-C-eps}), (\ref{eq:FT-phi-C}) and (\ref{eq:FT-F}), we find the field strength and dilaton change to 
\begin{eqnarray}
e^{\tilde{\phi}^\prime}\tilde{F}^\prime=-\frac{4}{R}\gamma_{\hat{0}\hat{1}\hat{2}\hat{3}\hat{4}}+\frac{4}{R}(\gamma_{\hat{0}\hat{1}\hat{2}\hat{3}\hat{4}}-i\gamma_{\hat{4}})=-i\frac{4}{R}\gamma_{\hat{4}},\quad \tilde{\phi}^\prime=-4\log(\frac{z}{R}),
\end{eqnarray}
where we have rewritten the equations in \cite{Berkovits:2008ic} in our notations. The NS-NS metric and B-field do not change under this transformation.
Writing by using the coordinate $r=\frac{R^2}{z}$, we can find that the resulting background coincides with \eqref{eq:TsTT4}. 
Then performing the bosonic T-duality tranform along the non-radial 
directions, we can obtain the NCAdS background 
\eqref{eq:NCAdS-background}. 

We give comment on this result as following: Note that the final dual background depends on the choice of Killing spinor in (\ref{eq:FT-rule}). 
However, no matter which set of Killing spinors we choose, the metric and B-field will not change under the fermionic T-duality transformation. Our transformation can be regarded as a simplification of the NCAdS background (or NCSYM). If the detail map from the observables in the NCAdS background \textcolor{blue}{\eqref{eq:NCAdS-background}} to the observables in (\ref{eq:TsTT4FT}) background is known, one can calculate the observables in (\ref{eq:TsTT4FT}) background easily. In the following sections, we show two examples of these observables to show the powerfulness of our result.

\section{The scattering amplitude in the NCAdS spacetime}\label{sec:Amplitude-NC}
We study the IR regime of the gluon scattering amplitude at strong coupling in NCSYM.
We follow the procedure in the original $AdS_5\times S^5$ spacetime case \cite{Alday:2007hr}, and study the open string scattering amplitude on D3-brane near horizon in NCAdS background (\ref{eq:NCAdS-background}). We simplify this problem by studying the object in the final dual background (\ref{eq:TsTT4FT}).

\subsection{Open string boundary condition before and after T-dual transformation}
Just as in the study of gluon scattering amplitude in \cite{Alday:2007hr}, we consider the Euclidean worldsheet.
The bosonic part of the worldsheet action on background (\ref{eq:NCAdS-background}) is
\bee
\frac{1}{4\pi\alpha'}\int d^{2}\sigma\sqrt{g}(g^{ab}G_{\mu\nu}+i\epsilon^{ab}B_{\mu\nu})\partial_{a}x^\mu\partial_b x^\nu.
\label{eq:ws-action-01}
\ee
The open strings on the D3-brane near horizon satisfy the boundary condition 
\begin{eqnarray}\label{eq:b.c.NC}
\big( G_{\mu\nu}\partial_{\sigma}x^{\nu}+ iB_{\mu\nu}\partial_{\tau}x^{\nu}\big)|_{\partial\Sigma}=0,
\end{eqnarray}
where $\partial\Sigma$ is the boundary of the worldsheet. This is mixed Neumann-dirichlet boundary conditions in the directions $x^0,\cdots,x^3$. In components this boundary condition can be written as
\begin{eqnarray}
0&=&G_{0\nu}\partial_{\sigma}x^{\nu}|_{\partial\Sigma}\\
0&=&G_{3\nu}\partial_{\sigma}x^{\nu}|_{\partial\Sigma}\\
0&=&G_{1\nu}\partial_{\sigma}x^{\nu}+iB_{12}\partial_{\tau}x^{2}|_{\partial\Sigma}\\
0&=&G_{2\nu}\partial_{\sigma}x^{\nu}+iB_{21}\partial_{\tau}x^{1}|_{\partial\Sigma}.
\end{eqnarray}

We then consider the bosonic T-duality transform along $x^1,x^2,x^3,x^0$ step by step. We will use (\ref{eq:T-dual-coord}) to study the transformation of the boundary conditions. We label the field $f$ after the T-duality transformation along direction $x^a$ as $f_{(a)}$, where $a=1,2,3,0$, $f_{(0)}=\tilde{f}$ is the field in background (\ref{eq:TsTT4FT}) \footnote{In this section, we only consider the field $f$ which is invariant under ferimionic T-dual transformation. This means the field in (\ref{eq:TsTT4}) and (\ref{eq:TsTT4FT}) are same.}.

We show the first step of T-duality transformation along direction $x^1$ for instance.
We start with the boundary condition:
\bee\label{eq:b.c.NC-0}
G_{\mu\nu}\partial_{\sigma}x^{\nu}|_{\partial\Sigma}=-i\epsilon^{\sigma\tau}B_{\mu\nu}\partial_{\tau}x^{\nu}|_{\partial\Sigma}.
\ee
Taking T-dual transformation along $x^{1}$ direction, (\ref{eq:T-dual-coord}) leads to
get
\bee
\partial_{\alpha}x_{(1)}^{1}&=&-i\epsilon_{\alpha\beta}\big(G_{11}\partial_{\beta}x^{1}+G_{1m}\partial_{\beta}x^{m}\big)-B_{1m}\partial_{\alpha}x^{m}\\
\partial_{\alpha}x_{(1)}^{2}&=&\partial_{\alpha}x^{2},~\partial_{\alpha}x_{(1)}^{3}=\partial_{\alpha}x^{3},~\partial_{\alpha}x_{(1)}^{0}=\partial_{\alpha}x^{0}\no
\ee
Resolving these equations, one obtains
\bee
\partial_{a}x^{1}&=&i\epsilon_{ba}G_{12(1)}\partial_{b}x_{(1)}^{2}+i\epsilon_{ba}G_{11(1)}\partial_{b}x_{(1)}^{1}\\
\partial_{\alpha}x^{2}&=&\partial_{\alpha}x_{(1)}^{2},~~\partial_{\alpha}x^{3}=\partial_{\alpha}x_{(1)}^{3},~~\partial_{\alpha}x^{0}=\partial_{\alpha}x_{(1)}^{0}\no
\ee
Then the boundary condition becomes
\bee
\partial_{\tau}x_{(1)}^{1}|_{\partial\Sigma}&=&0,~~\partial_{\sigma}x_{(1)}^{3}|_{\partial\Sigma}=0,\qquad \partial_{\sigma}x_{(1)}^{0}|_{\partial\Sigma}=0,\\
\partial_{\sigma}x_{(1)}^{2}|_{\partial\Sigma}&=&-\frac{\gamma R^{2}}{r^{2}}\big(G_{12(1)}\partial_{\sigma}x_{(1)}^{2}+G_{11(1)}\partial_{\sigma}x_{(1)}^{1}\big)_{\partial\Sigma}.
\ee
In the same way, we study the boundary condition after the T-duality transformation along $x^2, x^3,x^0$ in turn and 
find the boundary condition in the final dual coordinates 
becomes the simple Dirichlet condition
\bee\label{eq:b.c.after-T-dual}
\partial_{\tau}\tilde{x}^{\mu}|_{\partial\Sigma}&=&0,~~\mu=0,1,2,3.
\ee
Since we started with the D3-brane near horizon ($r\to \infty$) in background (\ref{eq:NCAdS-background}), the open strings are then fixed at the AdS boundary $(z=\frac{R^2}{r}\to 0)$ in background (\ref{eq:TsTT4FT}).

Note that the boundary condition of open strings does not change under 
the fermionic T-duality transformation.
Therefore the scattering amplitude on the D-brane in NCAdS spacetime is mapped 
to a T-dual open strings with Dirichlet boundary condition in the background 
(\ref{eq:TsTT4FT}). Furthermore since the T-duality transform does not change 
the boundary condition of radial direction, the open string has the Dirichlet 
condition along $z$ direction.

\subsection{Null Polygon Wilson loop}
We then study which object in background (\ref{eq:TsTT4FT}) dual to the scattering amplitude in background (\ref{eq:NCAdS-background}). Following the process in \cite{Alday:2007hr}, we consider the scattering of the open strings ending on the IR D3-brane in background (\ref{eq:NCAdS-background}). We insert the vertex operator with a momentum $k^\mu$ on the boundary of the worldsheet. At this boundary, $x^\mu$ thus carries a momentum $k^\mu$, i.e. the zero mode of field $x^{\mu}$. In background (\ref{eq:TsTT4FT}), this translates to the condition about the ``winding'' $\Delta x^\mu=x^\mu(\sigma=2\pi)-x^\mu(\sigma=0)=\int_{0}^{2\pi}d\sigma\partial_{\sigma}\tilde{x}^{\mu}$.   
From (\ref{eq:T-dual-coord}), we find that $\partial_{\sigma}\tilde{x}^{\mu}$ and $\partial_{\sigma}x^{\mu}$ 
are related with
\bee
\partial_{\sigma}\tilde{x}^{1}
&=&-i\epsilon_{\sigma\tau}G_{11}\partial_{\tau}x^{1}-B_{12}\partial_{\sigma}x^{2}\\
\partial_{\sigma}\tilde{x}^{2}&=&-i\epsilon_{\sigma\tau}G_{22}\partial_{\tau}x^{2}+B_{12}\partial_{\sigma}x^{1}\\
\partial_{\sigma}\tilde{x}^{3}&=&-i\epsilon_{\sigma\tau}G_{33(2)}\partial_{\tau}x^{3}\\
\partial_{\sigma}\tilde{x}^{0}&=&-i\epsilon_{\sigma\tau}G_{00(2)}\partial_{\tau}x^{0}.
\ee
 Then $\Delta \tilde{x}^{\mu}=\int_{0}^{2\pi}d\sigma\partial_{\sigma}\tilde{x}^{\mu}$ can be written as
\bee
\Delta \tilde{x}^{1}&=& 
k^{1}-B_{12}x^{2}|_{\sigma=0}^{2\pi}\\
\Delta \tilde{x}^{2}
&=& 
k^{2}+B_{12}x^{1}|_{\sigma=0}^{2\pi}\\
\Delta \tilde{x}^{3}&=&
k^{3}\\
\Delta \tilde{x}^{0}&=&
k^{0},
\ee
where $k^\mu=
k^\mu_{prop}\frac{R^2}{r^2}$ and $k^\mu_{prop}$ are the momentum carried by vertex operator and the proper momentum of string in (\ref{eq:NCAdS-background}) respectively. We then use the boundary condition (\ref{eq:b.c.NC-0}) to rewrite $\partial_\sigma x^\mu$ into $k^{\mu}$
\bee
\Delta \tilde{x}_{i}^{1}&=&
k_i^{1}-\frac{B_{12}B_{21}}{G_{11}G_{22}}k_i^1\\
\Delta \tilde{x}_{i}^{2}
&=&
k_i^{2}-\frac{B_{12}B_{21}}{G_{11}G_{22}}k_i^2\\
\Delta \tilde{x}_{i}^{3}&=&
k_i^{3}\\
\Delta \tilde{x}_{i}^{0}&=&
k_i^{0},
\ee
where $i$ is the label of the open strings.
Since momentum conservation ($\sum_ik_i^\mu=0$) of the scattering, the segments constructed by $\Delta \tilde{x}_{i}^{\mu}$ should be always closed.
Since we should take $r$ to $\infty$ (horizon), $B_{12}(\sim \frac{R^4}{r^4})$ is negligible comparing to $G_{11}\sim \frac{R^2}{r^2}$. We thus obtain
\bee\label{eq:Dx}
\Delta \tilde{x}_{i}^\mu=
k^{\mu}_i,
\ee
which shows that 
the segments are lightlike (null), because all the gluons are massless.
We thus obtain a null polygon Wilson loop at the boundary in (\ref{eq:TsTT4FT}).

We then consider the background (\ref{eq:TsTT4FT}), which is an $AdS_5\times S^5$ spacetime with constant B-field. 
The sigma action becomes
\begin{eqnarray}
S&=&\frac{1}{4\pi\alpha'}\int d^{2}\sigma\left(\partial^{a}\tilde{x}^{\mu}\partial_{a}\tilde{x}^{\nu}\tilde{G}_{\mu\nu}+i\tilde{B}_{\mu\nu}\epsilon^{ab}\partial_{a}\tilde{x}^{\mu}\partial_{b}\tilde{x}^{\nu}\right)\\
&=&\frac{1}{4\pi\alpha'}\left(\int d^{2}\sigma\left(\partial^{a}\tilde{x}^{\mu}\partial_{a}\tilde{x}^{\nu}\tilde{G}_{\mu\nu}\right)+i\int_{\partial\Sigma}\tilde{B}_{\mu\nu}\tilde{x}^{\mu}\partial_{t}\tilde{x}^{\nu}\right).\label{eq:ws-action} 
\end{eqnarray}
where $\partial_t$ is a derivative along the worldsheet boundary $\partial\Sigma$. Since the $\tilde{B}$-field is constant, it does not contribute to the e.o.m.
The second term of 
\eqref{eq:ws-action} depends only on the boundary and just gives the 
an overall dressing phase
\begin{eqnarray}
\Phi=\frac{i}{4\pi\alpha'}\int_{\partial\Sigma}d\sigma\gamma\big(\tilde{x}^{1}
\partial_{\sigma}\tilde{x}^{2}-\tilde{x}^{2}
\partial_{\sigma}\tilde{x}^{1}\big),
\end{eqnarray}
where the boundary condition (\ref{eq:b.c.after-T-dual}) is used. Since the worldsheet ends on the null polygon Wilson loop with segments $\Delta \tilde{x}_i^\mu=k^\mu_i$ at boundary, we can parametrize the boundary as
\bee
\tilde{x}^\mu=\sum_mk_m^\mu\theta(\sigma-\sigma_m),
\ee
where $m$ is the label of the segments of Wilson loop, $\sigma_m$ is the location of the cusp. We thus obtain
\bee
\Phi=\frac{i\gamma}{4\pi\alpha^\prime}\sum_{m<n}\big(k_m^1k_n^2-k_m^2k_n^1\big).
\ee
This result matches with the discussion from the viewpoint of NCSYM in \cite{Sever:2009xu}. 
One could study the $\alpha^\prime$ correction (or $\sqrt{\lambda}$) of (\ref{eq:ws-action}) by considering the fluctuation around the classical solution $\tilde{x}^\mu$ \cite{Kruczenski:2007cy}. Since the boundary of (\ref{eq:ws-action}) should be fixed, the effect of the constant B-field does not change. This means at any order of $\alpha'$ (or $\sqrt{\lambda}$), the B-field only contributes a phase factor $\Phi$. This is consistent with our expectation that the non-commutative super Yang-Mills theory reduce to the original $N=4$ super Yang-Mills theory at IR regime (long distance)\cite{Alday:2007hr}.

To close this section, let us show the similarities and differences to the discussion in \cite{Sever:2009xu}. The author in \cite{Sever:2009xu} studied the gluon scattering amplitude from gravity side by using a slight different background \footnote{In \cite{Sever:2009xu}, the background corresponds to a $(2,2)$ signature NCSYM.}, and argued that the scattering amplitude dual to the null polygon Wilson loop in the T-dual background. In the present paper, basing on the Buscher rules in appendix A, we studied how the boundary condition and the momentum of strings transform under the T-duality. This provides solid proof to show that the scattering amplitude in  background (\ref{eq:NCAdS-background}) corresponds to the null polygon Wilson loop in (\ref{eq:TsTT4FT}).
We also used the fermionic T-dual transformation to justify the complex background, which has not been discussed in \cite{Sever:2009xu}. Furthermore, we have also considered the fluctuation around the classical solution, and found the B-field contribute the same phase factor $\Phi$ at any order of $\alpha^\prime$ (or $\sqrt{\lambda}$) in the planar limit, which is consistent with the expectation in \cite{Sever:2009xu} from the viewpoint of field theory side.

\section{Open string solution in NCAdS background}\label{sec:solu-NCAdS}\label{sec:Open string Solution in NCAdS}
The AdS/CFT correspondence enables us to study the strong coupling gauge theory by using the classical solution in gravity side. 
In this section we study the classical open solution in NCAdS background (\ref{eq:NCAdS-background}), which is dual to the folded string \cite{Gubser:2002tv} solution in background  (\ref{eq:TsTT4FT}). 

If the string configuration does not depend on directions $x^1$ and $x^2$, the classical solution of this type string configuration should be described by the same classical solution \footnote{Some of those string configurations have been given by \cite{Kruczenski:2012aw}.} as in the original $AdS_5\times S^5$ spacetime. We will study non-trivial classical solution which depends on $x^1$ and $x^2$ directions.
However due to the appearance of B-field and the complicated metric in (\ref{eq:NCAdS-background}), it is not an easy work to construct this kind classical solution. Indeed in \cite{Maldacena:1999mh}, the authors studied the Wilson line in (\ref{eq:NCAdS-background}), and suggest that the {open} strings cannot be localized near the boundary in background (\ref{eq:NCAdS-background}).
This is one of the difficulties to study the non-commutative super Yang-Mills theory from gravity side. In this section, we show one procedure to solve this problem.

We start with the classical solution in background (\ref{eq:TsTT4FT}), and use the Buscher rule to find the dual classical solution in NCAdS spacetime (\ref{eq:NCAdS-background}). 
More precisely, we start with the folded string \cite{Gubser:2002tv} in original $AdS_5\times S^5$ spacetime, which plays an important role in the study of AdS/CFT correspondence.
Our calculation in this section can be regarded as the NCAdS spacetime version of the folded string studied in \cite{Kruczenski:2012aw}.

Following the usual notation for folded string in literature, we use the Lorentizian signature worldsheet action
\bee\label{eq:Action-Lorentizian}
S=-\frac{1}{4\pi\alpha'}\int d\tau d\sigma\big(\eta^{\alpha\beta}\partial_{\alpha}x^{\mu}\partial_{\beta}x^{\nu}G_{\mu\nu}-\epsilon^{\alpha\beta}B_{\mu\nu}\partial_{\alpha}x^{\mu}\partial_{\beta}x^{\nu}\big),
\ee
where $\epsilon^{\tau\sigma}=1$, and $\eta^{\alpha\beta}$ is the metric of worldsheet in Lorentization signature.
 Note that the background (\ref{eq:TsTT4FT}) does not change by varying the signature of worldsheet. However, the transformation of coordinate becomes (\ref{eq:Buscher-coordinate-Lorentz}).
{For the Euclidean signature worldsheet, the T-duality transformation will map the real solutions to the complex ones (see \cite{Dekel:2016oot} for recent developments). Here we are considering the Lorentizian signature, such that the T-duality will map the real solutions to the real ones.}

\subsection{Coordinates and folded string solutions in original AdS spacetime}
We first consider the folded string solution in the global coordinate of original $AdS_5\times S^5$ spacetime. To fix the notation, we should describe the coordinates that was used from now. The embedding coordinates of original $AdS_5\times S^5$ spacetime are defined by
\bee
ds_{_{AdS_{5}}}^{2}=dX_{M}dX^{M},\quad-X_{M}X^{M}=X_{-1}^{2}+X_{0}^{2}-X_{1}^{2}-X_{2}^{2}-X_{3}^{2}-X_{4}^{2}=1,
\ee
where we have set $R=1$ for simplicity.
The global coordinates$(t,\rho,\Omega_3)$ is given by
\bee
ds^{2}&=-\cosh^{2}\rho dt^{2}+d\rho^{2}+\sinh^{2}\rho(d\phi^{2}+\cos^{2}\phi d\theta_{1}^{2}+\sin^{2}\phi d\theta_{2}^{2}).
\ee
with 
\bee
X_{0}+iX_{-1}&=\cosh\rho e^{it},\quad X_{1}+iX_{2}=\sinh\rho\cos\phi e^{i\theta_{1}},\quad X_{3}+iX_{4}=\sinh\rho\sin\phi e^{i\theta_{2}}.
\ee
The Poincare coordinates are defined by
\bee
z&=\frac{1}{X_{-1}-X_{4}},\quad x^{0}=\frac{X_{0}}{X_{-1}-X_4},\quad x^{i}&=\frac{X^{i}}{X_{-1}-X_4},\quad i=1,2,3,
\ee
whose metric is given by
\bee
ds^{2}=\frac{dz^{2}+dx_{\mu}dx^{\mu}}{z^{2}}.
\ee

 The folded string is solved under the ansatz in global coordinate
of AdS$_{5}$ space
\begin{align*}
t & =\kappa\tau,\quad\theta_{1}=\kappa\omega\tau,\quad\phi=\theta_{2}=0.
\end{align*}
In the conformal gauge, the e.o.m. and Virasoro constrains become
\begin{align*}
\rho^{\prime\prime}+\kappa^{2}(\omega^{2}-1)\sinh\rho\cosh\rho & =0\\
\rho^{\prime2}-\kappa^{2}(\cosh^{2}\rho-\omega^{2}\sinh^{2}\rho) & =0.
\end{align*}
This is solved by \footnote{Our notation for  is different with the one used in \cite{Kruczenski:2012aw}. }
\begin{align}\label{eq:gloded-global}
\sinh(\rho)  =\frac{1}{\omega}\frac{\mbox{sn}(\kappa\omega\sigma|\frac{1}{\omega^{2}})}{\mbox{dn}(\kappa\omega\sigma|\frac{1}{\omega^{2}})},\quad
\cosh(\rho)  =\frac{1}{\mbox{dn}(\kappa\omega\sigma|\frac{1}{\omega^{2}})},
\end{align}
where $\mbox{sn}$ and $\mbox{dn}$ are the Jacobi Elliptic functions. See Appendix \ref{sec:Jacobi Elliptic function} for the definition.
Here we have focused on the case $\omega^2>1$. The folded string rotates with the angular velocity $\frac{d\theta_1}{dt}=\omega$. The radial coordinate $\rho(\sigma)$ varies in the range $(0, \mbox{arctanh}(\frac{1}{\omega}))$,which is fixed by the condition $\frac{d\rho}{d\sigma}|_{\rho\to \rho_0}=0$.  
Since the constant B-field does not effect the e.o.m. and Virasoro constraints, the classical folded string solution in original $AdS_5\times S^5$ spacetime is also a solution in background (\ref{eq:TsTT4FT}). 
Writing the solution (\ref{eq:gloded-global}) in the background (\ref{eq:TsTT4FT}), we obtain the classical folded string solution
\begin{align}
{z}&=\frac{\mbox{dn}(\kappa\omega\sigma|\frac{1}{\omega^{2}})}{\sin(\kappa\tau)},\quad
\tilde{x}^{0}=\frac{\cos(\kappa\tau)}{\sin(\kappa\tau)}\label{eq:large-spin-solu}\\
\tilde{x}^{1}&=\frac{1}{\omega}\mbox{sn}(\kappa\omega\sigma|\frac{1}{\omega^{2}})\frac{\cos(\kappa\omega\tau)}{\sin(\kappa\tau)},\quad \tilde{x}^{2}=\frac{1}{\omega}\mbox{sn}(\kappa\omega\sigma|\frac{1}{\omega^{2}})\frac{\sin(\kappa\omega\tau)}{\sin(\kappa\tau)}.\no
\end{align}

\subsection{Classical solution in NCAdS spacetime dual to folded string}
Our task is to find the corresponding T-dual(back) solution in background (\ref{eq:NCAdS-background}), which is the background before the T-dual transformation. By definition, $r$ can be obtained by
\bee
r&=&\frac{1}{\tilde{z}}=\frac{\sin(\kappa\tau)}{\mbox{dn}(\kappa\omega\sigma|\frac{1}{\omega^{2}})}.
\ee
 The Buscher rule for coordinates (\ref{eq:Buscher-coordinate-Lorentz}) leads to
\bee
\partial_{\beta}\tilde{x}^{1}&=&\epsilon_{\beta\alpha}\partial^{\alpha}x^{1}\frac{{\cal M}}{r^{2}}-\partial_{\beta}x^{2}\frac{{\cal M\gamma}}{r^{4}}\no\\
\partial_{\beta}\tilde{x}^{2}&=&\epsilon_{\beta\alpha}\partial^{\alpha}x^{2}\frac{1}{r^{2}}+\epsilon_{\beta\alpha}\eta^{\alpha\gamma}\Big(\epsilon_{\gamma\delta}\partial^{\delta}x^{1}\frac{{\cal M}\gamma}{r^{4}}-\partial_{\gamma}x^{2}\frac{{\cal M}\gamma^{2}}{r^{6}}\Big)\no\\
\partial_{\beta}\tilde{x}^{3}&=&\epsilon_{\beta\alpha}\partial^{\alpha}x^{3}\frac{1}{r^{2}}\label{eq:Buscher-rule-all-coordinate}\\
\partial_{\beta}\tilde{x}^{0}&=&-\epsilon_{\beta\alpha}\partial^{\alpha}x^{0}\frac{1}{r^{2}}.\no
\ee
where ${\cal M}^{-1}=1+\frac{\gamma^{2}R^{4}}{r^{4}}$, and $x^\mu$ means the coordinate in background (\ref{eq:NCAdS-background}).
Resolving $\partial_{\alpha}x^{\mu}$ by using $\partial_{\beta}\tilde{x}^{\mu}$,
we find
\begin{align}
\partial_{\sigma}x^{0} & =r^{2}\partial_{\tau}\tilde{x}^{0},\quad\partial_{\tau}x^{0}=r^{2}\partial_{\sigma}\tilde{x}^{0},\no\\
\partial_{\sigma}\big(x^{1}+ix^{2}\big) & =-i\gamma\partial_{\sigma}\big(\tilde{x}^{1}+i\tilde{x}^{2}\big)-r^{2}\partial_{\tau}\big(\tilde{x}^{1}+i\tilde{x}^{2}\big),\label{eq:tildex->x}\\
\partial_{\tau}\big(x^{1}+ix^{2}\big) & =-i\gamma\partial_{\tau}\big(\tilde{x}^{1}+i\tilde{x}^{2}\big)-r^{2}\partial_{\sigma}\big(\tilde{x}^{1}+i\tilde{x}^{2}\big).\no
\end{align}

Substituting (\ref{eq:large-spin-solu}) to (\ref{eq:tildex->x}), we obtain
\begin{align}
r&=\frac{\sin(\kappa\tau)}{\mbox{dn}(\kappa\omega\sigma|\frac{1}{\omega^{2}})},\quad
x^{0}=-\frac{\frac{\text{cn}\left(\omega\kappa\sigma\left|\frac{1}{\omega^{2}}\right.\right)\text{sn}\left(\omega\kappa\sigma\left|\frac{1}{\omega^{2}}\right.\right)}{\omega^{2}\text{dn}\left(\omega\kappa\sigma\left|\frac{1}{\omega^{2}}\right.\right)}-\mbox{E}\left(\text{am}\left(\omega\kappa\sigma\left|\frac{1}{\omega^{2}}\right.\right)|\frac{1}{\omega^{2}}\right)}{\left(\frac{1}{\omega^{2}}-1\right)\omega},\label{eq:solu:NCAdS}\\
x^{1}+ix^{2}&=-i\gamma\frac{e^{i\kappa \omega\tau}}{\sin(\kappa\tau)}\mbox{sn}(\kappa\omega\sigma|\frac{1}{\omega^{2}})-\frac{e^{i\kappa\omega\tau }\text{cn}\left(\omega\kappa\sigma\left|\frac{1}{\omega^{2}}\right.\right)(\cos(\kappa\tau)-i\omega\sin(\kappa\tau))}{(\omega^{2}-1)\text{dn}\left(\omega\kappa\sigma\left|\frac{1}{\omega^{2}}\right.\right)}\no.
\end{align}

Using Mathematica, one can check that the solution (\ref{eq:solu:NCAdS}) satisfy the e.o.m. in the background (\ref{eq:NCAdS-background}) 
\bee
&&r(x^{0})^{\prime\prime}-r\ddot{x}^{0}+2\dot{r}\dot{x}^{0}-2r^{\prime}(x^{0})^{\prime}=0\\
&&\big((x^{2})^{\prime\prime}-\ddot{x}^{2}\big)\left(\gamma^{2}r+r^{5}\right)+(x^{2})^{\prime}\left(2\gamma^{2}r^{\prime}-2r^{4}r^{\prime}\right)+\dot{x}^{2}\left(2r^{4}\dot{r}-2\gamma^{2}\dot{r})\right)+4\gamma r^{2}r^{\prime}\dot{x}^{1}-4\gamma r^{2}\dot{r}(x^{1})^{\prime}=0\no\\
&&\big((x^{1})^{\prime\prime}-\ddot{x}^{1}\big)\left(\gamma^{2}r+r^{5}\right)+(x^{1})^{\prime}\left(2\gamma^{2}r^{\prime}-2r^{4}r^{\prime}\right)+\dot{x}^{1}\left(2r^{4}\dot{r}-2\gamma^{2}\dot{r}\right)+4\gamma r^{2}\dot{r}(x^{2})^{\prime}-4\gamma r^{\prime}r^{2}\dot{x}^{2}=0\no\\
&&\big(r^{\prime\prime}-\ddot{r}\big)\left(\gamma^{4}r+2\gamma^{2}r^{5}+r^{9}\right)+\big(\dot{r}^{2}-(r^{\prime})^{2}\big)\left(2\gamma^{2}r^{4}+r^{8}+\gamma^{4}\right)+\big((\dot{x}^{0})^{2}-((x^{0})^{\prime})^{2}\big)\left(2\gamma^{2}r^{4}+r^{8}+\gamma^{4}\right)\no\\
&&+\big(((x^{1})^{\prime})^{2}-(\dot{x}^{1})^{2}\big)\left(r^{8}-\gamma^{2}r^{4}\right)+4\gamma r^{6}\big((x^{1})^{\prime}\dot{x}^{2}-\dot{x}^{1}(x^{2})^{\prime}\big)+\big((\dot{x}^{2})^{2}-((x^{2})^{\prime})^{2}\big)\left(\gamma^{2}r^{4}-r^{8}\right)=0.\no
\ee
and the Virasoro constraints
\bee
T_{\tau\tau}&=&T_{\sigma\sigma}=\frac{1}{2}\big(\dot{x}^{\mu}\dot{x}_{\mu}+(x^{\mu})'(x_{\mu})'\big)=0\\T_{\tau\sigma}&=&T_{\sigma\tau}=\dot{x}^{\mu}(x_{\mu})'=0,
\ee
where $\dot{}$ and $'$ mean the derivative about $\tau$ and $\sigma$ respectively\footnote{In background (\ref{eq:TsTT4FT}), the e.o.m. are 
\bee
r^{\prime\prime}+\frac{-(r^{\prime})^{2}+\dot{r}^{2}-(x^{0\prime})^{2}+(\dot{x}^{0})^{2}+(x^{1\prime})^{2}-(\dot{x}^{1})^{2}+(x^{2\prime})^{2}-(\dot{x}^{2})^{2}}{r}&=&\ddot{r},\no\\
(x^{0})^{\prime\prime}+\frac{2\dot{r}\dot{x}^{0}-2r^{\prime}x^{0\prime}}{r}&=&\ddot{x}^{0},\no\\
(x^{1})^{\prime\prime}+\frac{2\dot{r}\dot{x}^{1}-2r^{\prime}x^{1\prime}}{r}&=&\ddot{x}^{1}.
\ee
}. Therefore, (\ref{eq:solu:NCAdS}) is a classical solution in NCAdS background (\ref{eq:NCAdS-background}).
One could also follow the procedure in \cite{Kruczenski:2012aw}, and interchange $\tau$ and $\sigma$ to have an open string interpretation for this classical solution
\begin{align}
r&=\frac{\sin(\kappa\sigma)}{\mbox{dn}(\kappa\omega\tau|\frac{1}{\omega^{2}})},\quad x^{0}=-\frac{\frac{\text{cn}\left(\omega\kappa\tau\left|\frac{1}{\omega^{2}}\right.\right)\text{sn}\left(\omega\kappa\tau\left|\frac{1}{\omega^{2}}\right.\right)}{\omega^{2}\text{dn}\left(\omega\kappa\tau\left|\frac{1}{\omega^{2}}\right.\right)}-\mbox{E}\left(\text{am}\left(\omega\kappa\tau\left|\frac{1}{\omega^{2}}\right.\right)|\frac{1}{\omega^{2}}\right)}{\left(\frac{1}{\omega^{2}}-1\right)w},\label{eq:Open-solu-NCAdS}\\
x^{1}+ix^{2}&=-i\gamma\frac{e^{i\kappa \omega\sigma}}{\sin(\kappa\sigma)}\mbox{sn}(\kappa\omega\tau|\frac{1}{\omega^{2}})-\frac{e^{i\kappa \omega\sigma}\text{cn}\left(\omega\kappa\tau\left|\frac{1}{\omega^{2}}\right.\right)(\cos(\kappa\sigma)-i\omega\sin(\kappa\sigma))}{(\omega^{2}-1)\text{dn}\left(\omega\kappa\tau\left|\frac{1}{\omega^{2}}\right.\right)}.\no
\end{align}
 At the boundary of NCAdS background, i.e. $r=0$, $\sigma=0,\frac{\pi}{\kappa}$.
The worldsheet (\ref{eq:solu:NCAdS}) ends on two curves corresponding to $\sigma=0,\frac{\pi}{\kappa}$ respectively
\bee
x^{0}&=&-\frac{\frac{\text{cn}\left(\omega\kappa\tau\left|\frac{1}{\omega^{2}}\right.\right)\text{sn}\left(\omega\kappa\tau\left|\frac{1}{\omega^{2}}\right.\right)}{\omega^{2}\text{dn}\left(\omega\kappa\tau\left|\frac{1}{\omega^{2}}\right.\right)}-\mbox{E}\left(\text{am}\left(\omega\kappa\tau\left|\frac{1}{\omega^{2}}\right.\right)|\frac{1}{\omega^{2}}\right)}{\left(\frac{1}{\omega^{2}}-1\right)w}\no\\
x^{1}+ix^{2}&=&-i\gamma\frac{1}{\sin(\kappa\sigma=0)}\mbox{sn}(\kappa\omega\tau|\frac{1}{\omega^{2}})-\frac{\text{cn}\left(\omega\kappa\tau\left|\frac{1}{\omega^{2}}\right.\right)}{(w^{2}-1)\text{dn}\left(\omega\kappa\tau\left|\frac{1}{\omega^{2}}\right.\right)}.\label{eq:boundary-curve-1}
\ee
and
\bee
x^{0}&=-\frac{\frac{\text{cn}\left(\omega\kappa\tau\left|\frac{1}{\omega^{2}}\right.\right)\text{sn}\left(\omega\kappa\tau\left|\frac{1}{\omega^{2}}\right.\right)}{\omega^{2}\text{dn}\left(\omega\kappa\tau\left|\frac{1}{\omega^{2}}\right.\right)}-\mbox{E}\left(\text{am}\left(\omega\kappa\tau\left|\frac{1}{\omega^{2}}\right.\right)|\frac{1}{\omega^{2}}\right)}{\left(\frac{1}{\omega^{2}}-1\right)w},\no\\
x^{1}+ix^{2}&=-i\gamma\frac{e^{i\pi \omega}}{\sin(\kappa\sigma=\pi)}\mbox{sn}(\kappa\omega\tau|\frac{1}{\omega^{2}})+\frac{e^{i\pi \omega}\text{cn}\left(\omega\kappa\tau\left|\frac{1}{\omega^{2}}\right.\right)}{(\omega^{2}-1)\text{dn}\left(\omega\kappa\tau\left|\frac{1}{\omega^{2}}\right.\right)}.\label{eq:boundary-curve-2}
\ee
The two curves (\ref{eq:boundary-curve-1}) and (\ref{eq:boundary-curve-2}) at the boundary are related by the spatial rotation $-e^{i\kappa\omega\sigma}|_{\sigma=\frac{\pi}{\kappa}}$.
One may wonder weather it makes sense or not as $x^1+ix^2\to \infty$. It is easy to show $x^1+ix^2$ gives a finite value as it couple to the metric in (\ref{eq:NCAdS-background}). Note that (\ref{eq:solu:NCAdS}) can be regarded as a periodic open string  $\tau\in (0,2\pi)$ and $\sigma\in (0,\frac{\pi}{\kappa})$.

It is interesting to calculate the on-shell action which is equal to the minimal area of the surface ending on the Wilson loop. Substituting (\ref{eq:Open-solu-NCAdS}) to (\ref{eq:Action-Lorentizian}), we find
\begin{align}
-4\pi \alpha'S=&\int_0^{2\pi} d\tau\int_0^{\pi/\kappa} d\sigma 2\kappa^{2}\Big(\gamma\csc^{2}(\kappa\sigma)\text{cn}\left(\kappa\omega\tau\left|\frac{1}{\omega^{2}}\right.\right)\text{dn}\left(\kappa\omega\tau\left|\frac{1}{\omega^{2}}\right.\right)\text{sn}\left(\kappa\omega\tau\left|\frac{1}{\omega^{2}}\right.\right)\no\\&-\csc^{2}(\kappa\sigma)+\text{sn}\left(\kappa\omega\tau\left|\frac{1}{\omega^{2}}\right.\right)^{2}\Big)
\no\\
=&4 \pi ^2 \kappa  \omega^2-2 \pi \omega  \mbox{E}\left(\text{am}\left(2 \pi  \omega \kappa \left|\frac{1}{\omega^2}\right.\right)|\frac{1}{\omega^2}\right)+S_{div},
\end{align}
where $S_{div}$ is the divergent pieces given by
\bee
S_{div}=\left.\left(4 \pi  \kappa  \cot (\kappa  \sigma )+\frac{\gamma  \cot (\kappa  \sigma ) \left(\text{cn}\left(2 \pi  \omega \kappa \left|\frac{1}{\omega^2}\right.\right)^2-1\right)}{\omega}\right)\right|_{\sigma=0}^{\sigma=\pi/\kappa}.
\ee
The source of the divergence in the integration comes from the singular behavior of the integrand near the boundary. We perform the regularization by shifting the boundary from $r=0$ to $r=\epsilon$ with a small $\epsilon$. For a given $\tau$, the range of $\sigma$ change from $(0, \frac{\pi}{\kappa})$ to $(\frac{\arcsin\left(\epsilon  \text{dn}\left(\omega \kappa  \tau \left|\frac{1}{\omega^2}\right.\right)\right)}{\kappa} , \frac{\pi-\arcsin\left(\epsilon  \text{dn}\left(\omega \kappa  \tau \left|\frac{1}{\omega^2}\right.\right)\right)}{\kappa} )$.
We first integrate over $\sigma$ with a fixed $\tau$ and find
\begin{align}
-4\pi \alpha'S&=\int_0^{2\pi}d\tau 2 \kappa  \left(\frac{2 \gamma  \text{cn}\left( \kappa\omega \tau \left|\frac{1}{\omega^2}\right.\right) \text{sn}\left( \kappa\omega \tau \left|\frac{1}{\omega^2}\right.\right) \sqrt{1-\epsilon ^2 \text{dn}\left( \kappa\omega \tau \left|\frac{1}{\omega^2}\right.\right)^2}}{\epsilon }\right.\\
&\left.+\text{sn}\left( \kappa\omega \tau \left|\frac{1}{\omega^2}\right.\right)^2 \left(\pi -2 \sin ^{-1}\left(\epsilon  \text{dn}\left(\kappa\omega \tau \left|\frac{1}{\omega^2}\right.\right)\right)\right)-\frac{2 \sqrt{1-\epsilon ^2 \text{dn}\left(\kappa\omega \tau \left|\frac{1}{\omega^2}\right.\right)^2}}{\epsilon  \text{dn}\left(\kappa\omega \tau \left|\frac{1}{\omega^2}\right.\right)}\right)\no
\end{align}
Expanding the integrand by using $\epsilon$, one finds
\begin{align}
-4\pi \alpha'S&=\int_0^{2\pi}d\tau\frac{4 \gamma  \kappa  \text{cn}\left(\tau \omega \kappa \left|\frac{1}{\omega^2}\right.\right) \text{sn}\left(\tau \omega \kappa \left|\frac{1}{\omega^2}\right.\right)-\frac{4 \kappa }{\text{dn}\left(\kappa\omega\tau \left|\frac{1}{\omega^2}\right.\right)}}{\epsilon }+2 \pi  \kappa  \text{sn}\left(\kappa\omega\tau \left|\frac{1}{\omega^2}\right.\right)^2\\
&-2 \epsilon  \left(\kappa  \text{dn}\left(\kappa\omega\tau \left|\frac{1}{\omega^2}\right.\right) \left(\gamma  \text{cn}\left(\kappa\omega\tau \left|\frac{1}{\omega^2}\right.\right) \text{dn}\left(\kappa\omega\tau \left|\frac{1}{\omega^2}\right.\right) \text{sn}\left(\kappa\omega\tau \left|\frac{1}{\omega^2}\right.\right)+2 \text{sn}\left(\kappa\omega\tau \left|\frac{1}{\omega^2}\right.\right)^2-1\right)\right)\no\\
&+O\left(\epsilon ^3\right).\no
\end{align}
Then integrating over $\tau$, one obtains
\begin{align}
-4\pi \alpha'S&=\frac{4\gamma\omega\left[1-\text{dn}\left(2\pi\kappa\omega\left|\frac{1}{\omega^{2}}\right.\right)\right]-\frac{4}{\sqrt{\omega^{2}-1}}\arccos\left(\frac{\text{cn}\left(2\pi\kappa\omega\left|\frac{1}{\omega^{2}}\right.\right)}{\text{dn}\left(2\pi\kappa\omega\left|\frac{1}{\omega^{2}}\right.\right)}\right)}{\epsilon }\no\\
&+4 \pi ^2 \kappa  \omega^2-2 \pi  \omega \mbox{E}\left(\text{am}\left(2 \pi  \kappa\omega \left|\frac{1}{\omega^2}\right.\right)|\frac{1}{\omega^2}\right)\no\\
&+\epsilon  \left(\frac{2 \text{cn}\left(2 \pi  \kappa\omega \left|\frac{1}{\omega^2}\right.\right) \text{sn}\left(2 \pi   \kappa\omega \left|\frac{1}{\omega^2}\right.\right)}{\omega}+\frac{2}{3} \gamma \omega \text{dn}\left(2 \pi \kappa\omega \left|\frac{1}{\omega^2}\right.\right)^3-\frac{2 \gamma \omega}{3}\right).\label{eq:open-string-on-shell}
\end{align}
Note that the parameter $\gamma$ appears in the divergence term, but does not appear in the $\epsilon^0$ term. 
The divergence in (\ref{eq:open-string-on-shell}) has the form of $\frac{1}{\epsilon}$, which is the same form as the one of quark-antiquark Wilson loop case but different {from the divergent behavior ($\log\frac{1}{\epsilon}$) in cusp Wilson loop case  \cite{Drukker:1999zq}.} The divergence in (\ref{eq:open-string-on-shell}) can be interpreted as a self energy of heavy quark pairs in NCSYM.

The subtracted on-shell action is
\begin{align}
(-4\pi \alpha'S)_{reg}&=(1+\epsilon\frac{\partial}{\partial\epsilon})(-4\pi \alpha'S)
=4 \pi ^2 \kappa  \omega^2-2 \pi  \omega \mbox{E}\left(\text{am}\left(2 \pi  \kappa\omega \left|\frac{1}{\omega^2}\right.\right)|\frac{1}{\omega^2}\right),\no
\end{align}
where we have dropped the $\epsilon$ term. Here we just make use of minimal substraction to regular the action. We notice that the subtracted on-shell action does not depend on the parameter $\gamma$.

\section{Conclusions and Discussions}
In this paper, we have performed certain bosonic T-duality and fermionic T-duality transformations on the NCAdS background, and found the final dual background is the usual $AdS_5\times S^5$ background but with a constant NS-NS B-field depending on the non-commutative parameter. Our transformation can be regarded as the simplification of the NCAdS background, which is very useful to study the physics in non-commutative super Yang-Mills theory. As application, we have studied the gluon scattering amplitudes and Wilson loop in the NCSYM holographically by using the simplified final dual background.
In the final dual background, we found the worldsheet ending on the null polygon Wilson loop dual to the gluon scattering amplitudes in the NCSYM theory, which extends the scattering amplitude/Wilson loop duality for the NCSYM. We found that the non-commutative deformation will contribute to the gluon scattering amplitude as overall dressing phase phase factor shown in section \ref{sec:Amplitude-NC}, which is valid even for the finite ${\lambda}$. Furthermore, motivated by the relation between closed and open strings \cite{Kruczenski:2012aw}, we started with the folded string in the final dual background, and constructed the periodic open string (Wilson loop) solution in the NCAdS background. We have also calculated the on-shell action of the open string solution, which describes the minimal area of the  ending on the Wilson loop. The divergence of the on-shell action appears in the form of $\frac{1}{\epsilon}$, where $\epsilon$ is the small regularization parameter. We also noticed that the subtracted on-shell action does not depend on the non-commutative parameter.

On the other hand, an open-closed string map for the TsT deformation has been argued in a series of papers \cite{vanTongeren:2015soa,vanTongeren:2015uha,Osten:2016dvf,vanTongeren:2016eeb,Araujo:2017jkb,Araujo:2017jap,Sakamoto:2017cpu, Bakhmatov:2018apn}. Using the open-closed string map on the TsT deformation background, one finds the open string metric and coupling go back to the original background. The information about TsT deformation only appear in the non-commutative parameter. The series T-duality transformation in our present work also map the TsT deformation background to the original background but with a inverse radial coordinate in the metric. It is an interesting problem to explore the relations between our works and the open-closed string map.

It would be interesting to study the bosonic and fermionic T-duality transformations on other type of TsT deformation of the $AdS_5\times S^5$ background, which may lead to a simplification in the same way.
We also would like to construct the fermionic T duality of deformed ABJM theory \cite{Imeroni:2008cr}. Unlike $AdS_5\times S_5$, the holographic background of ABJM is not self-dual background by fermionic T-dual, it will be very interesting and highly non-trivial to holographically investigate the corresponding string solutions which correspond to gluon scattering amplitude, Wilson Loop operator and integrability structures \cite{He:2013hxd} of anomalous dimension of local operators in various deformed ABJM theories in the future. 

\section*{Acknowledgement}
We would like to thank K. Ito, H. Nakajima and J. Sakamoto for the useful discussions. We also thank E.~O.~Colgain and M. M. Sheikh-Jabbari for very useful comments on this manuscript.
S.H. are grateful to the Mainz institute for Theoretical Physics (MITP) and Les Houches workshop on structures in local quantum field at Ecole de Physique des Houches for its hospitality and its partial supports during the completion of this work. S.H. is supported from Max-Planck fellowship in Germany, the German-Israeli Foundation for Scientific Research and Development. The work of H.S. are supported in part by JSPS Research Fellowship for Young Scientists, from the Japan Ministry of Education, Culture, Sports, Science and Technology

\appendix
\section{Bosonic T-dual transformation}\label{sec:BT-dual}
We summarize the Buscher rule that is used in this paper. One can also see \cite{Alvarez:1994dn,Imeroni:2008cr} for review.
We start with the Euclidean worldsheet action as following
\bee\label{eq:WS-Ploy}
S=\frac{1}{4\pi \alpha'}\int d^2\sigma (\sqrt{h}h^{ab}g_{mn}+i\epsilon^{ab}b_{mn})\partial_{a}x^m\partial_bx^n.
\ee
and assume that the background field $g_{mn}$ and $b_{mn}$ are invariant under the shift isometry:
	\begin{eqnarray}
		x^1\to x^1+c,~~x^{\hat{m}}\to x^{\hat{m}},
	\end{eqnarray}
	where $c$ is a constant and $\hat{m}\neq 1$. We denote the field $f$ ($f=g_{mn}, b_{mn}, \phi, x^m$) after the T-dual transformation as $f'$. We do the T dual operation along $x^1$ and then the Buscher rule can be summarized as
	\begin{eqnarray}
		{g}_{11}'&=&\frac{1}{g_{11}},\quad{g}'_{1i}=\frac{b_{1i}}{g_{11}},\quad{g}'_{ij}=g_{ij}-\frac{g_{1i}g_{1j}-b_{1i}b_{1j}}{g_{11}}\no\\
        {b}_{1i}'&=&\frac{g_{1i}}{g_{11}},\quad{b}'_{ij}=b_{ij}-\frac{g_{1i}b_{1j}-b_{1i}g_{1j}}{g_{11}},\\
\phi'&=&\phi-\frac{1}{2}\log |g_{11}|.\no
	\end{eqnarray}
The coordinates transform as
\bee\label{eq:T-dual-coord}
\partial_a{x'}^1&=&-i\epsilon_{ab}[g_{11}\partial_b{x}^{1}+g_{1\hat{m}}\partial_bx^{\hat{m}}]-b_{1\hat{m}}\partial_ax^{\hat{m}},~~~~{x'}^{\hat{m}}=x^{\hat{m}}.
\ee
We then summarize the transformation of RR fields. Given a $p-$form $\omega_p$, we decompose it as
	\bee
	\omega_p=\bar{\omega}_p+\omega_{p[y]}\wedge dy,
	\ee
	where $\bar{\omega}_p=\frac{1}{p!}\omega_{\alpha_1\cdots\alpha_p}dx^{\alpha_1}\wedge \cdots \wedge dx^{\alpha_p}$ does not contain $dy$ component. $\omega_{p[y]}$ is a $(p-1)$ form as $(\omega_{p[y]})_{\alpha_1\cdots\alpha_{p-1}y}$.
	For convention, we define the two one-form fields $j$ and $b$ as:
	\bee
	j&=&\frac{G_{\alpha y}}{G_{yy}}dx^{\alpha},\quad b=B_{[y]}+dy.
	\ee
	The T-duality rules for the R-R potential are then given by
	\bee
	C_{p}'&=&C_{p+1[y]}+\bar{C}_{p-1}\wedge b+C_{p-1[y]}\wedge b\wedge j.
	\ee
	In the modified field strength ${\cal F}_p=F_p+H\wedge C_{p-3}$, they change as
	\bee\label{eq:T-dual_RR}
	{\cal F}_{p}'&=&{\cal F}_{p+1[y]}+\bar{{\cal F}}_{p-1}\wedge b+{\cal F}_{p-1[y]}\wedge b\wedge j
	\ee
	where we used $db=H_{[y]}$.
	
	Sometimes, it is conventional to use Lorentizian worldsheet action, e.g. the study of GKP string:
	\bee
	S=-\frac{1}{4\pi\alpha'}\int d\tau d\sigma[\eta^{\alpha\beta}\partial_{a}x^{m}\partial_{\beta}x^{n}g_{mn}-\epsilon^{\alpha\beta}b_{mn}\partial_{\alpha}x^{m}\partial_{\beta}x^{n}].
	\ee
	In this case, (\ref{eq:T-dual-coord}) becomes
	\bee\label{eq:Buscher-coordinate-Lorentz}
\epsilon^{\alpha\beta}\partial_{\beta}{x}^{\prime 1}=\eta^{\alpha\beta}\partial_{\beta}x^{m}G_{1m}-\epsilon^{\alpha\beta}\partial_{\beta}x^{m}B_{1m}.
	\ee
	
    \section{Killing spinor equations in original $AdS_5\times S^5$ spacetime}\label{sec:KSE-two}
    We consider the Killing spinor equations in the original spacetime
\bee
ds^{2}&=&\frac{R^{2}}{r^{2}}[dx^{2}+dr^{2}]+R^{2}ds_{5}^{2}=\frac{R^{2}}{r^{2}}[dx^{2}+\sum^6_{j=1}dy_jdy_j],\\
F_{5}&=&-4R^4\left(\frac{1}{r^{5}}dt\wedge dx_{1}\wedge dx_{2}\wedge dx_{3}\wedge dr+\omega_{S^5}\right),
\ee
where $|y|=r$. Then the Killing spinor equations become
\bee
\delta\psi_{\hat{M}}&=&{e_{\hat{M}}}^{M}\nabla_{M}\epsilon+\frac{1}{2}\frac{1}{R}\gamma^{\hat{0}\hat{1}\hat{2}\hat{3}\hat{4}}\gamma_{\hat{M}}\hat{\epsilon}=0,\\
\delta\hat{\psi}_{\hat{M}}&=&{e_{\hat{M}}}^{M}\nabla_{M}\hat{\epsilon}-\frac{1}{2}\frac{1}{R}\gamma^{\hat{0}\hat{1}\hat{2}\hat{3}\hat{4}}\gamma_{\hat{M}}\epsilon=0.
\ee
We are mainly interested on the Killing spinor in \cite{Berkovits:2008ic}, where $\hat{\epsilon}=i\epsilon$ and $\epsilon$ is independent of the the coordinates $M=x^{0,1,2,3}$. This leads to the relation
\bee
i\gamma^{\hat{0}\hat{1}\hat{2}\hat{3}\hat{4}}\epsilon&=\gamma_{\hat{4}}\epsilon.
\ee
Then the Killing spinor equations become
\bee\label{eq:Killingeq-AdS}
{e_{\hat{j}}}^{j}\nabla_{j}\epsilon-\frac{1}{2R}\gamma_{\hat{j}}\gamma_{\hat{4}}\epsilon&=0,
\ee
where $j$ the coordinate of the remaining 6D part ($R\times S^5$). Since the remaining 6D part does not transform under our deformation. This leads to the Killing spinors (\ref{eq:KS-in-BM}) used in \cite{Berkovits:2008ic}. It is easy to find (\ref{eq:Killingeq-AdS}) has the same form as with (\ref{eq:Killingeq}). This is the reason why we choose the same $\epsilon$ as \cite{Berkovits:2008ic} to perform fermionic T-dual transformation in sec.2.
The difference between our Killing spinor and the chosen one in \cite{Berkovits:2008ic} is the signature of $\hat{\epsilon}$.

\section{Elliptic functions}\label{sec:Jacobi Elliptic function}
In this appendix, we summarize the Elliptic Integrals and Jacobi Elliptic functions that were used in sec.\ref{sec:Open string Solution in NCAdS}. We main follow the notations in \cite{Abramowitz-Stegun}. The elliptic integrals $\mbox{E}$ and $F$ are defined by:
\beeq
\mbox{E}(\phi|m)=\int_{0}^{\phi}\sqrt{1-m\sin^{2}\theta}d\theta,\quad \mbox{F}(\phi|m)=\int_{0}^{\phi}\frac{d\theta}{\sqrt{1-m\sin^{2}\theta}}.
\eeq
Using $\phi$ in $u=\mbox{F}(\phi|m)$, we define the Jacobi Elliptic functions by
\bee
\mbox{sn}(u|m)=\sin\phi,\quad\mbox{cn}(u|m)=\cos\phi,\quad\mbox{dn}(u|m)=\sqrt{1-m\sin^{2}\phi},\quad \phi=\mbox{am}(u|m).
\ee

	\end{document}